\documentclass{ws-ijmpb}

\begin{document}

\markboth{F. D. Klironomos and A. T. Dorsey}
{Tunneling current in bilayer electron system}

\catchline{}{}{}{}{}

\title{Tunneling between bilayer quantum Hall structures in a strong magnetic field}

\author{F. D. Klironomos}
\address{Department of Physics, University of Florida, P.O. Box 118440\\ Gainesville,
Florida 32611-8440, USA\\fkliron@phys.ufl.edu}

\author{A. T. Dorsey}
\address{Department of Physics, University of Florida, P.O. Box 118440\\ Gainesville,
Florida 32611-8440, USA\\dorsey@phys.ufl.edu}
\maketitle

\begin{history}
\received{23/06/2004}
\revised{16/08/2004}
\end{history}

\begin{abstract}
We calculate the tunneling current in a quantum Hall bilayer system in the 
strong magnetic field limit. We model the bilayer electron system as two Wigner
crystals coupled through interlayer Coulomb interactions, treated in the
continuum limit. We generalized the Johansson and Kinaret (JK) model\cite{jk}
and were able to study the effect of the low energy out-of-phase magnetophonon modes
produced as a result of tunneling events. We find the same scaling behavior of the
tunneling current peak with the magnetic field as found by JK but were able to 
find the tunneling current scaling behavior with interlayer distance as well.
\end{abstract}

\keywords{quantum Hall bilayer;two dimensional electron system}

\section{Introduction}
A two-dimensional bilayer electron system under the presence of a strong perpendicular magnetic field exhibits a
Coulomb barrier peak in the tunneling current in transport experiments\cite{eisenstein}. This peak is pronounced
in high density samples and is attributed to the highly correlated state of the electron liquid
where the cyclotron motion frustrates the ability of local electrons to respond to a tunneling event. A necessary
amount of energy of the order of $e^2/4\pi\epsilon\langle a\rangle$, where $\langle a\rangle$ is the average
interelectron distance, has to be provided to the system to collectively respond to tunneling.

Our model describes this tunneling behavior and is based on the Johansson and Kinaret (JK) model\cite{jk} which
assumes that the two layers of electrons are in commensurate Wigner crystal states.  A tunneling event disrupts
the Wigner crystals producing collective fluctuations (magnetophonons) that dissipate the energy of the tunneling 
electron and smooth-out the local charge-defect associated with it. We work in the continuum limit treating the 
whole system as an elastic medium. We should stress out that the actual state of the system is not a Wigner crystal 
but a rather highly correlated liquid state. Our work differs from the JK model in the fact that we have added 
interlayer Coulomb interaction, correlating the two layers, and additionally we were able to produce analytic 
results as well as numerical ones. Our results are summarized in Fig.~(\ref{fig1}).

\section{Bilayer theory}
\begin{figure}[th]
\centerline{\psfig{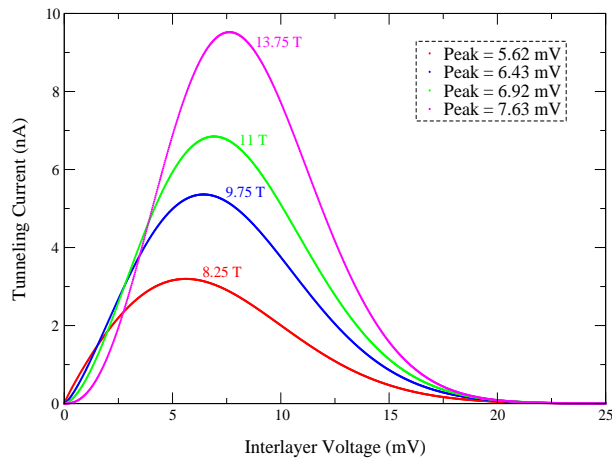}}
\vspace*{8pt}
\caption{Analytic results for the tunneling current of a two-dimensional bilayer system for different
magnetic field values. The legend shows the corresponding peak bias voltage.}
\label{fig1}
\end{figure}

We assume that each layer of electrons is a two-dimensional elastic medium interacting via Coulomb interaction (intralayer).
Each electron has a displacement associated with its equilibrium position in the Wigner crystal which becomes the displacement
field in the continuum limit. The magnetic field couples the transverse and longitudinal modes into gapless magnetophonon and 
gapped (by the cyclotron frequency) magnetoplasmon modes. These correspond to the internal modes of each layer when they are 
far apart (JK model). When the layers are brought within proximity the interlayer Coulomb interaction will introduce an 
additional coupling of those four modes producing two in-phase and two out-of-phase modes. The in-phase modes resemble single
layer modes but the out-of-phase modes are gapped by the energy cost to shear the commensurate Wigner crystals. This 
energy cost is associated with the penalty the short range part of the interlayer Coulomb interaction introduces to an 
out-of-phase motion of the two Wigner crystals. Additionally, a tunneling electron couples only to out-of-phase modes.
This happens because during tunneling the positive charge defect the electron leaves behind has to be ``covered-up"
while the receiving layer has to create a place for the incoming electron by ``opening-up" resulting in a relative 
out-of-phase motion of the two crystals.  The total Hamiltonian for the bilayer system assumes the form,
\begin{eqnarray}
H&=&H_0+H_T^-+H_T^+\nonumber\\
&=&\left[\epsilon_A+i\sum_sM_{sA}(a_s^\dagger-a_s)\right]c_A^\dagger c_A+\left[\epsilon_B+i\sum_sM_{sB}(a_s^\dagger-a_s)\right]
c_B^\dagger c_B \nonumber\\
&+&\sum_s\hbar\Omega_s a_s^\dagger a_s +Tc_A^\dagger c_B+ Tc_B^\dagger c_A,
\end{eqnarray}
where $\epsilon_A$, $\epsilon_B$ are the corresponding Madelung energies for the two Wigner crystals, $a_s^\dagger$, $a_s$ 
are the creation and annihilation operators for the collective electron modes while $c^\dagger$, $c$ are the tunneling 
electron's creation and annihilation operators for the two layers. The ground state of this Hamiltonian is a bath of collective
modes coupled to an external electron present in the system. Tunneling is an independent single particle process where the 
tunneling matrix element $T$ is assumed to be real and calculated as in JK\cite{jk}. The tunneling current for different 
bias voltage $V$ is given by
\begin{equation}
I(V)=\frac{e}{\hbar^2}\int_{-\infty}^{+\infty}\!\!dte^{ieVt/\hbar}\langle [H_T^-(t),H_T^+(0)]\rangle,
\end{equation}
We calculate the zero-temperature limit for the correlation function, since the experimental temperature range is $0.1$K
 $\sim$ 10$^{-5}$ eV while the applied bias voltage is in the mV range. Additionally we concentrate on the magnetophonon 
modes only that are responsible for the low bias Coulomb barrier peak. We find that the correlation function 
$\langle H_T^-(t)H_T^+(0)\rangle\sim C(t)$ can be Fourier transformed and obeys an integral equation of the form
\begin{equation}
\omega C(\omega)=\int_0^1\!\!dxf(x)\omega(x)C(\omega-\omega(x)),
\label{C}
\end{equation}
where $\omega(x)$ is the magnetophonon dispersion relation and $f(x)$ is a weight function. Equation (\ref{C}) has an
approximate large frequency solution of the form
\begin{equation}
C(\omega)\sim\exp\left\{-\frac{(\omega-c_1)^2}{2c_2}\right\},
\end{equation}
where $\sqrt{c_2}\sim 1/a_0^2\sqrt{B}$ for layers far apart and $c_2\sim d/a_0^3\sqrt{B}$ for layers close together, where
$d$ is the interlayer distance and $a_0$ the Wigner crystal lattice parameter. Based on that we propose a trial solution 
of the form
\begin{equation}
C(\omega)=N\omega^re^{-\lambda\omega^2},
\end{equation}
where $N$, $r$, $\lambda$ are self-consistently evaluated from the properties of $C(t)$ namely,
\begin{eqnarray}
\int_0^\infty\!\!d\omega C(\omega)=2\pi\\
\int_0^\infty\!\!d\omega\omega C(\omega)=2\pi c_1\\
\int_0^\infty\!\!d\omega\omega^2 C(\omega)=2\pi(c_2+c_1^2).
\end{eqnarray}
Our solutions are presented in Fig.~(\ref{fig1}) for different applied magnetic field values. For all our model parameters
we have used the experimental system reported in Ref.~\refcite{eisenstein}. For the magnetophonon gap value we extracted it
from time-dependent Hartree Fock calculations of the magnetophonon dispersion relation for the bilayer system\cite{cote}.
We numerically integrated Eq. (\ref{C}) to verify the validity of our analytic solution with very satisfactory results.

\section{Conclusions}
We calculated the tunneling current curve for a bilayer quantum Hall system under the presence of a strong perpendicular
magnetic field and reproduced the Coulomb barrier peak experimentally found. Our model assumed that the two dimensional
electron systems were in Wigner crystal states and calculated the eigenmodes in the continuum approximation where the
two layers are treated as elastic media. We found that by introducing an interlayer Coulomb interaction the internal modes
of the system couple to in-phase and out-of-phase ones. The latter acquire a gap in the magnetophonon branch associated
with the short-range interlayer correlations built into the system by the Coulomb force. This magnetophonon branch provides
the necessary collective mode coupling for tunneling electrons to dissipate their energy, and for the system to relax the 
charge defect a tunneling event creates. We found the same scaling behavior JK report for the tunneling curve but
additionally were able to extract the dependence on interlayer distance as well.

\section*{Acknowledgments}
We would like to thank S. M. Girvin for valuable discussion on the bilayers and Ren\'e C\^ot\'e and collaborators for providing
the magnetophonon dispersion curves. This work was supported by NSF DMR-9978547.

\appendix


\begin{thebibliography}{0}

\bibitem{jk}P.\ Johansson and J.\ M.\ Kinaret, Phys.\ Rev.\ {\bf B50}, 4671 (1994).

\bibitem{eisenstein}J.\ P.\ Eisenstein, L.\ N.\ Pfeiffer, and K.\ W.\ West Phys.\ Rev.\ Lett.\ {\bf 69}, 3804 (1992).

\bibitem{landau}L.\ D.\ Landau and E.\ M.\ Lifshitz, in {\it Theory of Elasticity}
(Butterworth--Heinemann, 3rd Edition), p.\ 9.

\bibitem{cote}Ren\'e C\^ot\'e {\it et al.}, private communication. 

\end{thebibliography}
\end{document}